\documentclass[letterpaper, 10 pt, conference]{ieeeconf}  

\IEEEoverridecommandlockouts                              
\usepackage{fancyhdr}
\usepackage{indentfirst}
\usepackage{amsmath}
\usepackage{amssymb}
\usepackage{graphicx}
\usepackage{color}
\usepackage{latexsym}
\usepackage{amsfonts}
\usepackage{amssymb,amsmath,amsfonts}
\usepackage{amsmath,mathrsfs,bm,url,times}

\newtheorem{theorem}{Theorem}
\newtheorem{lemma}{Lemma}

\newtheorem{definition}{Definition}

\newtheorem{remark}{Remark}

\title{\LARGE \bf
Self-Triggered Control for Multi-Agent Systems with\\ Quantized Communication or Sensing
}

\author{Xinlei Yi, Jieqiang Wei and Karl H. Johansson
\thanks{This work was supported by the
Knut and Alice Wallenberg Foundation, the  Swedish Foundation for Strategic Research, and the Swedish Research Council.}
\thanks{All the authors are with the ACCESS Linnaeus Centre, Electrical Engineering, KTH Royal Institute of Technology, 100 44, Stockholm, Sweden,
        {\tt\small \{xinleiy, jieqiang, kallej\}@kth.se}.}%
}

\begin{document}

\maketitle
\thispagestyle{empty}
\pagestyle{empty}

\begin{abstract}

The consensus problem for multi-agent systems with quantized communication
or sensing is considered. Centralized and distributed self-triggered rules
are proposed to reduce the overall need of communication and system
updates. It is proved that these self-triggered rules realize consensus
exponentially if the network topologies have a spanning tree and the
quantization function is uniform. Numerical simulations are provided to show
the effectiveness of the theoretical results.

\end{abstract}

\section{INTRODUCTION}

In the past decade, distributed cooperative control for multi-agent systems, particularly the consensus problem, has gained much attention and significant progress has been achieved, e.g., \cite{Saber2004}--\cite{Xiao2008}. Almost all studies assume that the information can be continuously transmitted between agents with infinite precision. In practice, such an idealized assumption is often unrealistic, so information transmission should to be considered in the analysis and design of consensus protocols \cite{You2011}.

There are two main approaches to handle the communication limitation: event-triggered and quantized control. In event-triggered (and self-triggered) control the control input is piecewise constant and transmission happens at discrete events \cite{Dimos2012}--\cite{Liu2015}. For instance, \cite{Dimos2012} provided event-triggered and self-triggered protocols in both centralized and distributed formulations for multi-agent systems with undirected graph topology; \cite{Liu2015} proposed a  self-triggered protocol for multi-agent systems with switching topologies. Other authors considered systems with quantized sensor measurements and  control inputs \cite{Dimos2010}--\cite{Guo2013}.

The authors of the papers \cite{Hu2012}--\cite{Garcia2013b} combined event-triggered control with quantized communication. For example, \cite{Garcia2013a} considered model-based event-triggered control for systems with quantization and time-varying network delays; \cite{Garcia2013b} presented decentralised event-triggered control in multi-agent systems with quantized communication.

When considering event-triggered control in multi-agent systems with quantized communication or sensing, some aspects should be paid special attention to. Firstly, the notion of the solution should be clarified since in some cases the classic or hybrid solutions may not exist. For instance, \cite{Ceragioli2011} and \cite{Guo2013} used the concept of Filippov solution when they considered quantized sensing. Secondly, the Zeno behavior must be excluded \cite{Joh1999}. Thirdly, the need of continuous state access for neighbors should be avoided. In \cite{Garcia2013b}, which is a key motivation for the present paper, the authors did not explicitly discuss the first aspect and used periodic sampling to exclude the Zeno behavior. They did not give any accurate upper bound of the sampling time, which restricts the application of the results.

Inspired by \cite{Xiao2008} and \cite{Liu2015}, we propose centralized and distributed self-triggered rules for multi-agent systems with quantized communication
or sensing. Under these rules, the existence of a unique trajectory of the system is guaranteed and the frequency of communication and system updating is reduced. The main contribution of the paper is to show that the trajectory exponentially converges to practical consensus set.  It is shown that continuously monitoring of the triggering condition can also be avoided. An important aspect of this paper is that the weakest fixed interaction topology is considered, namely, a directed graph containing a spanning tree. The proposed self-triggered rules are easy to implement in the sense that triggering times of each agent are only related to its in-degree.

The rest of this paper is organized as follows: Section \ref{sec2} introduces the preliminaries; Section \ref{sec3} discusses self-triggered consensus with quantized communication; Section \ref{sec4} treats instead self-triggered consensus with quantized sensing; simulations are given in Section \ref{sec5}; and the paper is concluded in Section \ref{sec6}.
\section{PRELIMINARIES}\label{sec2}

In this section we will review some results on algebraic graph theory \cite{Die}-\cite{Rah} and stochastic matrices \cite{Liu2011}-\cite{Paz1967}.

\subsection{Algebraic Graph Theory}
For a matrix $A \in\mathbb{R}^{n\times n}$,  the element at the $i$-th row and $j$-th column is denoted as $a_{ij}$; and denote $\overline{diag}(A)=A-diag([a_{11},\cdots,a_{nn}])$.

For a {\it (weighted) directed graph} (or digraph) $\mathcal G=(\mathcal V,\mathcal E, A)$ with $n$ agents (vertices or nodes), the set of
agents $\mathcal V =\{v_1,\dots,v_n\}$, set of links (edges) $\mathcal E
\subseteq \mathcal V \times \mathcal V$, and the {\it (weighted) adjacency matrix}
$A =(a_{ij})$ with nonnegative adjacency elements $a_{ij}$. A link
of $\mathcal G$ is denoted by $e(i,j)=(v_i, v_j)\in \mathcal E$ if there is
a directed link from agent $v_j$ to agent $v_i$ with weight $a_{ij}>0$, i.e. agent $v_j$ can send information to agent $v_i$ while the opposite direction transmission might not exist or with different weight $a_{ji}$. It is assumed that $a_{ii}=0$ for all $i\in \mathcal I$, where $\mathcal I=\{1,\dots,n\}$. Let $N^{in}_i=\{v_j\in \mathcal V\mid a_{ij}>0\}$ and $deg^{in}(v_i)=\sum\limits_{j=1}^{n}a_{ij}$ denotes the in-neighbors and in-degree of agent $v_i$, respectively. The degree matrix of digraph $\mathcal G$ is defined as $D=diag([deg^{in}(v_1), \cdots, deg^{in}(v_n)])$. The {\it (weighted) Laplacian matrix} is defined as $L=D-A$. A directed path from agent $v_0$ to agent $v_k$ is a directed graph with distinct agents $v_0,\dots,v_k$ and links $e(i+1,i),~i=0,\dots,k$.

\begin{definition}
We say a directed graph $\mathcal G$ has a spanning tree if there exists at least
one agent $v_{i_{0}}$ such that for any other agent $v_{j}$, there exits a
directed path from $v_{i_{0}}$ to $v_{j}$.
\end{definition}

Obviously, there is a one-to-one correspondence between a graph and its adjacency matrix or its Laplacian matrix. In the following, for the sake of simplicity in presentation, sometimes we don't explicitly distinguish a graph from its adjacency matrix or Laplacian matrix, i.e., when we say a matrix has some graphic properties, we mean that these properties are held by the graph corresponding to this matrix. 

\subsection{Stochastic Matrix}

A matrix $A =(a_{ij})$ is called a {\it nonnegative matrix} if $a_{ij}\ge0$ for all $i,j$, and $A$ is called a {\it stochastic matrix} if $A$ is square, nonnegative and $\sum_{j}a_{ij}=1$ for each $i$. A stochastic matrix $A$ is called {\it scrambling} if, for any $i$ and $j$, there exists $k$ such that both $a_{ik}$ and $a_{jk}$ are positive. Moreover, given a nonnegative matrix $A$ and $\delta>0$, the $\delta$-matrix of $A$, which is denoted as $A^{\delta}$, and its element at $i$-th row and $j$-th column, $a^\delta_{ij}$, is
\begin{align}
a^{\delta}_{ij}=
\begin{cases}
\delta,~a_{ij}\ge\delta\\
0,~a_{ij}<\delta
\end{cases}
\end{align}

If $A^{\delta}$ has a spanning tree, we say $A$ contains a $\delta$-spanning tree. Similarly, if $A^{\delta}$ is scrambling, we say $A$ is $\delta$-scrambling.

A nonnegative matrix $A$ is called a {\it stochastic indecomposable and aperiodic} (SIA) matrix if it is a stochastic matrix and there exists a column vector $v$ such that $\lim_{k\rightarrow\infty}A^k=\mathbf{1}v^{\top}$, where $\mathbf{1}$ is the $n$-vector containing only ones. For two $n$-dimension stochastic matrices $A$ and $B$, they are said to be of the same {\it type}, denoted by $A\sim B$, if they have zero elements and positive elements in the same places. Let $\mathbf{Ty}(n)$ denotes the number of different types of all SIA matrices in $\mathbb{R}^{n\times n}$, which is a finite number for given $n$. For two matrices $A$ and $B$ of the same dimension, we write $A\ge B$ if $A-B$ is a nonnegative matrix. Throughout this paper, we use $\prod_{i=1}^{k}A_i=A_kA_{k-1}\cdots A_1$ to denote the left product of matrices.

Here, we introduce some lemmas that will be used later.

From Corollary 5.7 in \cite{Liu2011}, we have
\begin{lemma}\label{lem1}
For a set of $n\times n$ stochastic matrices $\{A_1,A_2,\dots,A_{n-1}\}$, if there exists $\delta>0$ and $\delta'>0$ such that $A_k\ge\delta I$ and $A_k$ contains a $\delta'$-spanning tree for all $k=1,2,\dots,n-1$, then there exists $\delta''\in(0,\min\{\delta,\delta'\})$, such that $\prod_{k=1}^{n-1}A_k$ is $\delta''$-scrambling.
\end{lemma}

From Lemma 6 in \cite{Xiao2008}, we have
\begin{lemma}\label{lem2}
Let $A_1,A_2,\dots,A_k$ be $n\times n$ matrices with the property that for any $1\le k_1<k_2\le k$, $\prod_{i=k_1}^{k_2-1}A_i^{\delta}$ is SIA, where $\delta>0$ is a constant, then $\prod_{i=1}^{k}A_i$ is $\delta^{k}$-scrambling for any $k>\mathbf{Ty}(n)$.
\end{lemma}

\begin{definition}(\cite{Wu2006})
For a real matrix $A=(a_{ij})$, define the ergodicity coefficient $\mu(A)=\min_{i,j}\sum_{k}\min\{a_{ik},a_{jk}\}$ and its Hajnal diameter $\Delta(A)=\max_{i,j}\sum_{k}\max\{0,a_{ik}-a_{jk}\}$.
\end{definition}
\begin{remark}\label{remark1}
Obviously, if $A$ is a stochastic matrix, then $0\le\mu(A),\Delta(A)\le1$. Moreover, if $A$ is $\delta$-scrambling for some $\delta>0$, then $\mu(A)\ge\delta$.
\end{remark}

\begin{lemma}\label{lem3}
(\cite{Haj1958,Paz1967}) If $A$ and $B$ are stochastic matrices, then
$
\Delta(AB)\le(1-\mu(A))\Delta(B)
$.
\end{lemma}
\begin{lemma}\label{lem4}
(\cite{Liu2011}) For a vector $x=[x_1,\cdots,x_n]^{\top}\in\mathbb{R}^n$, define $d(x)=\max_i\{x_i\}-\min_{i}\{x_i\}$. For an $n\times n$ stochastic matrix $A$, and $x\in\mathbb{R}^n$, then
$
d(Ax)\le \Delta(A)d(x)\le \Delta(A)\sqrt{2}\|x\|
$.
\end{lemma}
\begin{remark}
It is straightforward to see that for any $x,y\in\mathbb{R}^n$, $d(x+y)\le d(x)+d(y)$.
\end{remark}

\section{Self-Triggered Control with Quantized Communication}\label{sec3}

We consider a set of $n$ agents that are modelled as a single integrator:
\begin{align}
\dot{x}_i(t)=u_i(t),~~i\in\mathcal I\label{system}
\end{align}
where $x_i(t)\in\mathbb{R}$ is the state and $u_i(t)\in\mathbb{R}$ is the input of agent $v_i$, respectively.

In many practical scenarios, each agent cannot access the state of the system with infinite precision. Instead, the state variables have to be quantized in order to be represented by a finite number of bits to be used in processor operations and to be transmitted over a digital communication channel. 

In this section, each agent has a self-triggered control input based on the latest quantized states of its in-neighbours. Denoting the {\em triggering time sequence} for agent $v_j$ as the increasing time sequence $\{t_{k}^{j}\}_{k=1}^{\infty}$, the control input is given as
\begin{align}
u_i=\sum_{j\in N^{in}_i}l_{ij}[q(x_i(t^i_{k_i(t)}))-q(x_j(t^j_{k_j(t)}))]\label{input}
\end{align}
where $k_{i}(t)=\arg\max_{k}\{t^{i}_{k}\le t\}$, $q:\mathbb{R}\rightarrow\mathbb{R}$ is a quantizer. In this paper, we consider the following {\em uniform} quantizer:
\begin{align}
|q_u(a)-a|\le \delta_u,~\forall a\in\mathbb{R}\label{uquan}
\end{align}

\begin{remark}
Compared with other papers, we do not need any additional assumptions about the quantizing function. For example, we do not need the quantizer to be an odd or monotonic function. However, at this moment we do not incorporate logarithmic quantizers as they do not satisfy (\ref{uquan}).
\end{remark}

\subsection{Centralized Triggering}\label{qcc}

In this subsection, we consider centralized self-triggered control, i.e., all agents simultaneously trigger at every triggering time. In this case, the triggering time sequence can be denoted as $t_1,t_2,\dots$. From (\ref{system}) and (\ref{input}), we get:
\begin{align}
\dot{x}(t)=-Lq(x(t_k)),~~t\in(t_k,t_{k+1}]\label{systemc}
\end{align}
where $x(t)=[x_{1}(t),\cdots,x_{n}(t)]^{\top}$ and $q(v)=[q(v_{1}),\cdots,q(v_{n})]^{\top}$ for any $v\in\mathbb{R}^{n}$.

Here we give a rule to determine the triggering time sequence such that all agents converge to practical consensus.
\begin{theorem}\label{thm1}
Assume the communication graph is directed, and contains a $\delta$-spanning tree with $\delta>0$. Given the first triggering time $t_1$, use the following self-triggered rule to find $t_{2},t_3\dots$ for
known $t_k$, choose an arbitrary $t_{k+1}\in(t_k+t_l,t_k+ t_u)$, where $t_l=\frac{\delta'}{L_{max}}$, $t_u=\frac{1-\delta'}{L_{max}}$, $\delta'\in(0,\frac{1}{2})$ and $L_{max}=\max_{i}l_{ii}$.
Then the trajectory of (\ref{systemc}) exponentially converges to the consensus set
$
\{x\in\mathbb{R}^{n}|d(x)\le C_1\delta_u\}
$,
where $C_1=[\frac{n-1}{\delta''}+1]4(1-\delta')$ and $\delta''\in(0,\min\{\delta',\frac{\delta'\delta}{L_{max}}\})$.
\end{theorem}

\begin{proof} From the self-triggered rule, for any given $t_1$, the system can arbitrarily choose $t_2\in[t_1+t_l,t_1+t_u]$ for every agent. Similarly, after $t_k$ has been chosen, the system can arbitrarily choose $t_{k+1}\in[t_k+t_l,t_k+t_u]$ for every agent. Then, in the interval $(t_k,t_{k+1}]$, the only solution\footnote{Different from other papers that consider quantization, here we can explicitly write out the unique solution.} to (\ref{systemc}) is
$
x(t)=x(t_k)-(t-t_k)Lq(x(t_k))
$.
Particularly, we have
\begin{align*}
x(t_{k+1})&=x(t_k)-\Delta t_kLq(x(t_k))\\
&=A_kx(t_k)-\Delta t_kLQ(x(t_k))
\end{align*}
where $\Delta t_k=t_{k+1}-t_k$, $A_k=I-\Delta t_kL$ and $Q(v)=q(v)-v$ for any $v\in\mathbb{R}^{n}$. Then
\begin{align*}
x(t_2)=&A_1x(t_1)-\Delta t_1LQ(x(t_1))\\
x(t_3)=&A_2x(t_2)-\Delta t_2LQ(x(t_2))\\
=&A_2A_1x(t_1)-A_2\Delta t_1LQ(x(t_1))-\Delta t_2LQ(x(t_2))\\
\vdots~&\\
x(t_{k+1})=&A_kx(t_k)-\Delta t_kLQ(x(t_k))\\
=&\prod_{i=1}^{k}A_ix(t_1)-\sum_{i=1}^{k-1}\prod_{j=i+1}^{k}A_j\Delta t_iLQ(x(t_i))\\
&-\Delta t_kLQ(x(t_k))
\end{align*}

Obviously, for every $k$, $A_k$ is a stochastic matrix; $A_k$ has a $\frac{\delta\delta'}{L_{max}}$-spanning tree since $L$ has a $\delta$-spanning tree and $\Delta t_k\ge \frac{\delta'}{L_{max}}$; $A_k>\delta'I$ since $[A_k]_{ii}=1-\Delta t_kL_{ii}\ge1-\Delta t_kL_{max}\ge\delta'$. Then, from Lemma \ref{lem1}, for any positive integer $k_0$, we know that $\prod_{i=k_0}^{k_0+n-2}A_i$ is $\delta''$-scrambling for some $0<\delta''<\delta'<\frac{1}{2}$.

Then from Remark \ref{remark1}, Lemma \ref{lem3} and Lemma \ref{lem4}, we have
\begin{align*}
d(x(t_{k+1}))<&(1-\delta'')^{n(k)}d(x(t_1))\\
&+\sum_{i=0}^{n(k-1)}(n-1)(1-\delta'')^i
t_u4L_{max}\delta_u
\end{align*}
where $n(k)=\lfloor \frac{k}{n-1}\rfloor$.
Thus
\begin{align*}
\overline{\lim_{k\to +\infty}}d(x(t_{k+1}))\le \frac{4(n-1)(1-\delta')}{\delta''}\delta_u
\end{align*}
For any $t>t_1$, there exists a positive integer $k$ such that $t\in(t_k,t_{k+1}]$. Then, we have
\begin{align*}
x(t)=[I-(t-t_k)L]x(t_k)-(t-t_k)LQ(x(t_k))
\end{align*}
Thus,
$
d(x(t))\le d(x(t_k))+4(1-\delta')\delta_u
$.
Hence
\begin{align*}
\overline{\lim_{t\rightarrow+\infty}}d(x(t))\le&
\overline{\lim_{k\rightarrow+\infty}}d(x(t_{k}))+4(1-\delta')\delta_u
\le
C_1\delta_u
\end{align*}
The proof is completed.
\end{proof}

\subsection{Distributed Triggering}

In this subsection, we consider distributed self-triggered control. In contrast to the centralized triggering where all agents trigger at the same time, each agent can now freely choose its own triggering times no matter when other agents trigger.
Here we extend Theorem \ref{thm1} to distributed such a distributed setup.
\begin{theorem}\label{thm2}
Assume the communication graph is directed, and contains a $\delta$-spanning tree with $\delta>0$. For each agent $v_i$, given the first triggering time $t^i_1$, use the following self-triggered rule to find $t^i_2,\dots,t^i_{k},\dots$ for
known $t^i_k$, choose an arbitrary $t^i_{k+1}\in(t^i_k+t^i_l,t^i_k+t^i_u)$,
where $t^i_l=\frac{\delta_i}{l_{ii}}$, $t^i_u=\frac{1-\delta_i}{l_{ii}}$ and $\delta_i\in(0,\frac{1}{2})$. Then the trajectory of (\ref{system}) with input (\ref{input}) exponentially converges  to the consensus set
$
\{x\in\mathbb{R}^{n}|d(x)\le C_2\delta_u\}
$,
where $C_2$ is a positive constant which can be determined by $\delta,\delta_1,\dots,\delta_n$.
\end{theorem}

\begin{proof} (a) (This proof is inspired by \cite{Xiao2008} and \cite{Liu2015}.) We say the system triggers at time $t$ if there exists at least one agent triggers at this time. Let $\{t_1,t_2,\dots\}$ denotes the system's triggering time sequence. Obviously, this is a strictly increasing sequence. For simplicity, denote $\Delta t^i_k=t^i_{k+1}-t^i_{k}$ and $\Delta t_k=t_{k+1}-t_k$. We first point out the following fact:
\begin{lemma}\label{lem5}
For any agent $v_i$ and positive integer $k$, the number of triggers occurred during $(t^i_{k},t^i_{k+1}]$ is no more than $\tau_1=(\lceil\frac{t_{max}}{t_{min}}\rceil+1)(n-1)$,  where $t_{min}=\min_{i}\{t^1_l,\dots,t^n_l\}$ and $t_{max}=\max_{i}\{t^1_u,\dots,t^n_u\}$. Moreover, for any positive integer $k$, every agent triggers at least once during $(t_k,t_{k+\tau_2}]$, where $\tau_2=(\lceil\frac{t_{max}}{t_{min}}\rceil+1)n$.
\end{lemma}
The proof of this lemma can be found in \cite{Liu2015}.

Let $y(t_k)=[y_1(t_k),y_2(t_k),\cdots,y_n(t_k)]^{\top}$ with $y_i(t_k)=x_i(t^i_{k_i(t_k)})$, then, we can rewrite (\ref{system}) and (\ref{input}) as
\begin{align}
\dot{x}_i(t)=-\sum_{j=1}^{n}l_{ij}q(y_j(t_k)),~t\in(t_k,t_{k+1}]\label{systemd}
\end{align}

Now we consider the evolution of $y(t_k)$.
If agent $v_i$ does not trigger at time $t_{k+1}$, then $t^{i}_{k_i(t_{k+1})}=t^i_{k_i(t_k)}$. Thus
\begin{align}
y_i(t_{k+1})=y_i(t_k)\label{solutiony1}
\end{align}
If agent $v_i$ triggers at time $t_{k+1}$, then $t^{i}_{k_i(t_{k+1})}=t_{k+1}$. Assume $t^i_{k_i(t_k)}=t_{k-d_{ik}}$ be the last update of agent $v_i$ before $t_{k+1}$, where integer $d_{ik}\ge0$ is the number of triggers which are triggered by other agents between $(t^i_{k_i(t_k)},t^{i}_{k_i(t_{k+1})})$. Then,
$
y_i(t_k)=y_i(t_{k-1})=\dots=y_i(t_{k-d_{ik}})
$.

Noting $(t^i_{k_i(t_k)},t^{i}_{k_i(t_{k+1})}]=\bigcup_{m=k-d_{ik}}^{k}(t_m,t_{m+1}]$ and (\ref{systemd}), we can conclude that there exists a unique solution to (\ref{system}). Then
\begin{align*}
&y_i(t_{k+1})
=x_i(t^{i}_{k_i(t_{k+1})})=x_i(t^{i}_{k_i(t_{k})})
+\int_{t^{i}_{k_i(t_{k})}}^{t^{i}_{k_i(t_{k+1})}}\dot{x}_i(t)dt\nonumber\\
=&y_i(t_{k-d_{ik}})+\sum_{m=k-d_{ik}}^k\int_{t_m}^{t_{m+1}}\dot{x}_i(t)dt\nonumber\\
\end{align*}
\begin{align}
=&y_i(t_k)-\sum_{m=k-d_{ik}}^k\Delta t_m\sum_{j=1}^{n}l_{ij}q(y_j(t_m))\nonumber\\
=&y_i(t_k)-\sum_{m=0}^{d_{ik}}\Delta t_{m+k-d_{ik}}\sum_{j=1}^{n}l_{ij}q(y_j(t_{m+k-d_{ik}}))\nonumber\\
=&y_i(t_k)-\sum_{m=0}^{d_{ik}}\Delta t_{m+k-d_{ik}}l_{ii}q(y_i(t_{m+k-d_{ik}}))\nonumber\\
&-\sum_{m=0}^{d_{ik}}\Delta t_{m+k-d_{ik}}\sum_{j\neq i}^{n}l_{ij}q(y_j(t_{m+k-d_{ik}}))\nonumber\\
=&y_i(t_k)-\sum_{m=0}^{d_{ik}}\Delta t_{m+k-d_{ik}}l_{ii}q(y_i(t_{k}))\nonumber\\
&-\sum_{m=0}^{d_{ik}}\Delta t_{m+k-d_{ik}}\sum_{j\neq i}^{n}l_{ij}q(y_j(t_{m+k-d_{ik}}))\nonumber\\
=&y_i(t_k)-\Delta t^i_{k_i(t_k)}l_{ii}q(y_i(t_{k}))\nonumber\\
&-\sum_{m=0}^{d_{ik}}\Delta t_{m+k-d_{ik}}\sum_{j\neq i}^{n}l_{ij}q(y_j(t_{m+k-d_{ik}}))\label{solutiony2}
\end{align}
If agent $v_i$ triggers at time $t_{k+1}$, then let $a^0_{ii}(k)=1-\Delta t^i_{k_i(t_k)}l_{ii}$, $b^0_{ii}(k)=-\Delta t^i_{k_i(t_k)}l_{ii}$, $a^m_{ii}(k)=b^m_{ii}(k)=0$ for $m=1,2,\dots,\tau_1$, $a^m_{ij}(k)=b^m_{ij}(k)=-\Delta t_{k-m}l_{ij}$ for $i\neq j$ and $m=0,1,2,\dots,d_{ik}$, and $a^m_{ij}(k)=b^m_{ij}(k)=0$ for $i,j=1,\dots,n$ and $m=d_{ik}+1,\dots,\tau_1$. Otherwise, let $a^m_{ij}(k)=b^m_{ij}(k)=0$ for all $i,j,m$ except $a^0_{ii}(k)=1$. Obviously,
\begin{align}
&a^0_{ii}(k)\ge 1-(1-\delta_i)=\delta_i\ge\delta_{min}\label{property1}\\
&-(1-\delta_{min})\le-(1-\delta_i)\le b^0_{ii}(k)\le-\delta_i\le-\delta_{min}\\
&\sum_{m=0}^{\tau_1}\sum_{j=1}^{n}a^m_{ij}(k)=1,\sum_{m=0}^{\tau_1}\sum_{j=1}^{n}b^m_{ij}(k)=0,
a^m_{ij}(k)\ge0\label{property2}
\end{align}
where $\delta_{min}=\min\{\delta_1,\dots,\delta_n\}$.

Then we can uniformly rewrite (\ref{solutiony1}) and (\ref{solutiony2}) as
\begin{align}
&y_i(t_{k+1})=\sum_{m=0}^{\tau_1}\sum_{j=1}^{n}a^m_{ij}(k)y_j(t_{k-m})\nonumber\\
+&\sum_{m=0}^{\tau_1}\sum_{j=1}^{n}b^m_{ij}(k)[q(y_j(t_{k-m}))-y_j(t_{k-m})]\label{solutiony3}
\end{align}
Denote $z(t_k)=[y(t_k)^{\top},y(t_{k-1})^{\top},\cdots,y(t_{k-\tau_1})^{\top}]^{\top}\in
\mathbb{R}^{n(\tau_1+1)}$, $A^m(k)=(a^m_{ij}(k))\in\mathbb{R}^{n\times n}$, $B^m(k)=(b^m_{ij}(k))\in\mathbb{R}^{n\times n}$,
\begin{eqnarray*}
C(k)=\left[\begin{array}{ccccc}
A^0(k)   &A^1(k)   &\cdots &A^{\tau_1-1}(k) &A^{\tau_1}(k)\\
I        &0        &\cdots &0               &0\\
0        &I        &\cdots &0               &0\\
\vdots   &\vdots   &\ddots &\vdots          &\vdots\\
0        &0        &\cdots &I               &0
\end{array}\right]
\end{eqnarray*}
and
\begin{eqnarray*}
D(k)=\left[\begin{array}{ccccc}
B^0(k)   &B^1(k)   &\cdots &B^{\tau_1-1}(k) &B^{\tau_1}(k)\\
0        &0        &\cdots &0               &0\\
0        &0        &\cdots &0               &0\\
\vdots   &\vdots   &\ddots &\vdots          &\vdots\\
0        &0        &\cdots &0               &0
\end{array}\right]
\end{eqnarray*}
From (\ref{property1}) and (\ref{property2}), we know that $C(k)$ is a stochastic matrix.
We can rewrite (\ref{solutiony3}) as
\begin{align}
z(t_{k+1})=C(k)z(t_k)+D(k)[q(z(t_k))-z(t_k)]
\end{align}

(b) Next, we will prove that there exists $\delta^C\in(0,1)$ such that for any $k_1>0$, $\prod_{k=k_1}^{K_0+k_1-1}C(k)$ is $\delta^C$ scrambling, where $K_0=(\mathbf{Ty}(n)+1)\tau_2$.

From (\ref{property1}) and (\ref{property2}), we know that $A^m(k)$ is a nonnegative matrix for any $m$ and $k$, and $A^0(k)\ge \delta_{min}I$. Hence, $\sum_{l=k}^{k+\tau_2}\sum_{m=0}^{\tau_1}A^m(l)\ge \delta_{min}I$.
Denote
\begin{eqnarray*}
M_0=\left[\begin{array}{ccccc}
I        &0        &\cdots &0               &0\\
I        &0        &\cdots &0               &0\\
0        &I        &\cdots &0               &0\\
\vdots   &\vdots   &\ddots &\vdots          &\vdots\\
0        &0        &\cdots &I               &0
\end{array}\right]
\end{eqnarray*}
and
$
C'(k)=\overline{diag}(C(k)-M_0)
$.
Then,
\begin{align}
C(k)\ge\delta_{min}M_0+C'(k)\ge\delta_{min}E(k)\label{ce}
\end{align}
where $E(k)=M_0+C'(k)$.

From Lemma \ref{lem5}, we know that, for any $k$, $\sum_{l=k}^{k+\tau_2}\sum_{m=0}^{\tau_1}A^m(l)\ge -\delta_{min}L$ since each agent triggers at least once during $(t_k,t_{k+\tau_2}]$. Hence, $\sum_{l=k+1}^{k+\tau_2}\sum_{m=0}^{\tau_1}A^m(l)$ has a $\delta_{min}\delta$-spanning tree. Thus, from Lemma 7 and its proof in \cite{Liu2015}, we know that there exists  $0<\delta_F^0<\delta_{min}\delta$ such that $F_k=\prod_{i=(k-1)\tau_2+1}^{k\tau_2}E(i)$ is $\delta_F$-SIA and has a $\delta_F$-spanning tree for any $\delta_F\in(0,\delta_F^0]$.
Here we choose a $\delta^F$ such that $0<\delta_F,(\delta_F)^{\frac{1}{\tau_2}}\le\delta_F^0<\delta_{min}\delta$.

For any $1\le k_1<k_2$, note
\begin{align*}
\prod_{k=k_1}^{k_2}F_k^{\delta_F}=&\prod_{k=k_1}^{k_2}\prod_{i=(k-1)\tau_2+1}^{k\tau_2}
[E(i)]^{[(\delta_F)^{\frac{1}{\tau_2}}]}\\
=&\prod_{i=(k_1-1)\tau_2+1}^{k_2\tau_2}
[E(i)]^{[(\delta_F)^{\frac{1}{\tau_2}}]}
\end{align*}
and the first block row sum of $\sum_{i=(k_1-1)\tau_2+1}^{k_2\tau_2}
[C'(i)]^{(\delta_F)^{\frac{1}{\tau_2}}}$ has a spanning tree since
$0<(\delta_F)^{\frac{1}{\tau_2}}\le\delta_F^0<\delta_{min}\delta$. Then from Lemma 7 and its proof in \cite{Liu2015}, we know that $\prod_{k=k_1}^{k_2}F_k^{\delta_F}$ is SIA.

Then, from Lemma \ref{lem2}, we know that $\prod_{k=k_1}^{\mathbf{Ty}(n)+k_1}F_k^{\delta_F}$ is $(\delta_F)^{(\mathbf{Ty}(n)+1)}$-scrambling. Hence, from  (\ref{ce}), we can conclude that $\prod_{k=(k_1-1)\tau_2+1}^{(\mathbf{Ty}(n)+k_1)\tau_2}C(k)$ is $\delta^{C}$-scrambling, where $0<\delta^C\le(\delta_F)^{(\mathbf{Ty}(n)+1)}(\delta_{min})^{K_0}$.

(c) Similar to the proof of Theorem \ref{thm1}, we can find the $C_2$ and complete the proof or this theorem.
\end{proof}
\begin{remark}
In both Theorems \ref{thm1} and \ref{thm2}, the evolutions of $x(t)$ obey $\xi^{\top}x(t)=\xi^{\top}x(0)$, where $\xi^{\top}L=0$.
\end{remark}

\begin{remark}
There is no Zeno behavior in the centralized and distributed self-triggered systems. Note that the triggering times are not dependent on the state, but the triggering rules are related only to the degree matrix.
\end{remark}
\section{Self-Triggered Control with Quantized Sensing}\label{sec4}

In this section, we consider the situation that, each agent $v_i$ discretely sense or measures the quantized value of the relative positions between its in-neighbors and itself. In other words, the only available information to compute the control inputs of each agent are the latest quantized measurements of the relative positions measured by itself:
\begin{align}
u_i(t)=\sum_{j\in N^{in}_i}a_{ij}q(x_j(t^i_{k_i(t)})-x_i(t^i_{k_i(t)}))\label{inputm}
\end{align}
\begin{remark}
Compared to (\ref{input}), the advantage of (\ref{inputm}) is that the input is not affected by other agents' triggering.
\end{remark}

\subsection{Centralized Triggering}\label{qmc}

In this subsection, we consider centralized self-triggered consensus rule and denote the triggering time sequence as $t_1,t_2,\dots$. Then, we get
\begin{align}
u_i(t)=\sum_{j\in N^{in}_i}a_{ij}q(x_j(t_k)-x_i(t_k)),~t\in(t_k,t_{k+1}]\label{inputmc}
\end{align}
Similar to Theorem \ref{thm1}, we have the following result.
\begin{theorem}\label{thm3}
Under the assumptions and self-triggered rule of Theorem \ref{thm1}, the trajectory of system (\ref{system}) with input (\ref{inputmc}) exponentially converges to the consensus set
$
\{x\in\mathbb{R}^{n}|d(x)\le C_3\delta_u\}
$,
where $C_3$ is a positive constant which can be determined by $\delta'$ and $\delta$.
\end{theorem}

\begin{proof} From the self-triggered rule in Theorem \ref{thm1}, for any given $t_1$, the system can arbitrarily choose $t_2\in[t_1+t_l,t_1+t_u]$ for every agent. Similarly, after $t_k$ has been chosen, the system can arbitrarily choose $t_{k+1}\in[t_k+t_l,t_k+t_u]$ for every agent. Then, in the interval $(t_k,t_{k+1}]$, the only solution to (\ref{system}) with input (\ref{inputmc}) is
\begin{align}
x_i(t)=x_i(t_k)+(t-t_k)\sum_{j=1}^{m}a_{ij}q(x_j(t_k)-x_i(t_k))\label{solutioncm}
\end{align}
Particularly, we have
\begin{align*}
x_i(t_{k+1})=x_i(t_k)+\Delta t_k\sum_{j=1}^{m}a_{ij}q(x_j(t_k)-x_i(t_k))
\end{align*}
Then,
$
x(t_{k+1})=A_kx(t_k)+\Delta t_kW(x(t_k))
$,
where $W(x(t_k))=[W_1(x(t_k)),W_2(x(t_k)),\cdots,W_n(x(t_k))]^{\top}$ and $W_i(x(t_k))=\sum_{j=1}^{m}a_{ij}[q(x_j(t_k)-x_i(t_k))-(x_j(t_k)-x_i(t_k))]$.
The proof follows similarly to the proof to Theorem \ref{thm1}.
\end{proof}

\subsection{Distributed Triggering}

In this subsection, we consider distributed self-triggered consensus rule. Similar to Theorem \ref{thm2}, we have
\begin{theorem}\label{thm4}
Under the assumptions and self-triggered rule of Theorem \ref{thm2},  the trajectory of (\ref{system}) with input (\ref{inputm}) exponentially converges  to the consensus set
$
\{x\in\mathbb{R}^{n}|d(x)\le C_4\delta_u\}
$,
where $C_4$  is a positive constant which can be determined by $\delta,\delta_1,\dots,\delta_n$.
\end{theorem}
\begin{proof}
We omit the proof since it is similar to the proof of Theorem \ref{thm2}.
\end{proof}

\section{SIMULATIONS}\label{sec5}
In this section, a numerical example is given to demonstrate the effectiveness of the presented results.

\begin{figure}[hbt]
\centering
\includegraphics[width=0.45\textwidth]{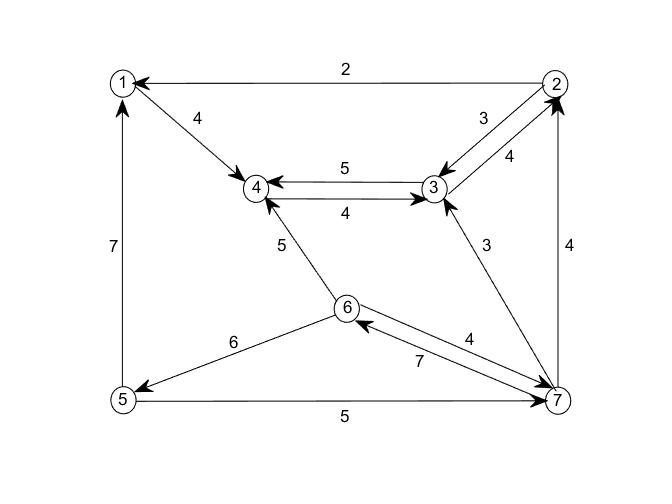}
\caption{The communication graph.}
\label{fig:1}
\end{figure}

Consider a network of seven agents with a directed reducible Laplacian matrix
\begin{eqnarray*}
L=\left[\begin{array}{rrrrrrr}9&-2&0&0&-7&0&0\\
0&8&-4&0&0&0&-4\\
0&-3&10&-4&0&0&-3\\
-4&0&-5&14&0&-5&0\\
0&0&0&0&6&-6&0\\
0&0&0&0&0&7&-7\\
0&0&0&0&-5&-4&9
\end{array}\right]
\end{eqnarray*}
which is described by the graph in Fig. \ref{fig:1}. The initial value of each agent is randomly selected within the interval $[-5,5]$ in our simulation and the next triggering time is randomly chosen from the permissible range using a uniform distribution. The uniform quantizing function used here is $q(v)=2k\delta_u$ if $v\in[(2k-1)\delta_u,(2k+1)\delta_u)$.

\begin{figure}[hbt]
\centering
\includegraphics[width=0.45\textwidth]{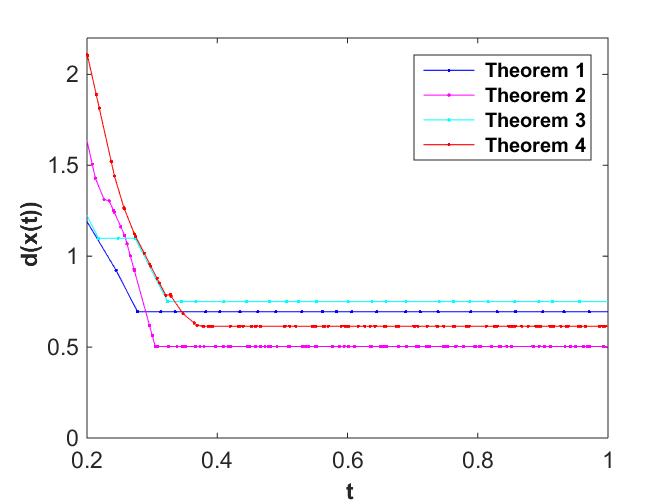}
\caption{The evolution of $d(x(t))$. The dots indicate the triggering times of each agent.}
\label{fig:2}
\end{figure}

Fig. \ref{fig:2} shows the evolution of $d(x(t))$ under the four self-triggered rules treated in Theorems \ref{thm1}-\ref{thm4} with $\delta_u=0.5$ and $\delta'=\delta_i=0.25$. In this simulation, it can be seen that under all self-triggering rules all agents converge to the consensus set with $C_1=C_2=C_3=C_4<2$.

\begin{figure}[hbt]
\centering
\includegraphics[width=0.45\textwidth]{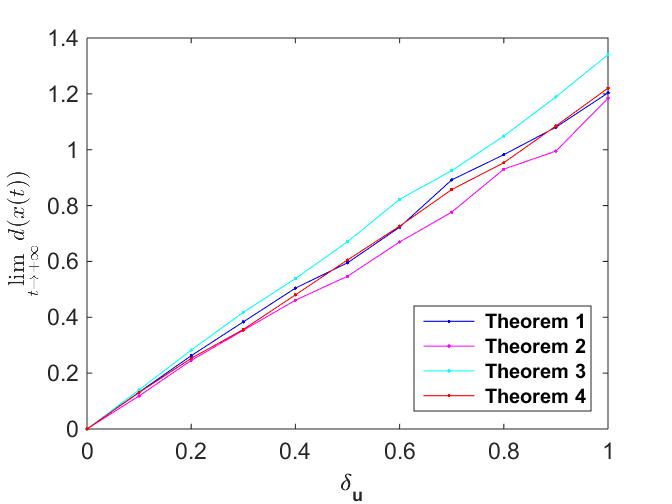}
\caption{The evolution of $\lim_{t\to +\infty}d(x(t))$ with different $\delta_u$.}
\label{fig:3}
\end{figure}

Let the quantizer parameter $\delta_u$ take different values. Fig. \ref{fig:3} illustrates $\lim_{t\to +\infty}d(x(t))$ under the four self-triggering rules for different $\delta_u$. The curves show the averages over 100 overlaps. As expected,  the smaller $\delta_u$, the smaller is the consensus set.

\section{CONCLUSIONS}\label{sec6}
In this paper, consensus problems for multi-agent systems defined on directed graphs under self-triggered control have been addressed. In order to reduce the overall need of communication and system updates, centralized and distributed self-triggered rules have been proposed in the situation that quantized information can only be transmitted, i.e., quantized communication, and the situation that each agent can sense only quantized value of the relative positions between neighbors, i.e., quantized sensing.
It has been shown that the trajectory of each agent exponentially converges to the consensus set if the directed graph containing a spanning tree. The triggering rules can be easily implemented since they are related only to the degree matrix. Interesting future directions include considering stochastically switching topologies and more precise expression of the consensus sets.

\end{document}